\documentclass[aps,pra,twocolumn,amssymb,amsmath]{revtex4-1}
\usepackage{amsmath}
\usepackage{amssymb}
\usepackage{latexsym}
\usepackage{enumitem}
\usepackage{revsymb}
\usepackage{epsfig}
\usepackage{color}

\begin{document}

\title{Conservation laws and the foundations of quantum mechanics}

\author{Yakir Aharonov$^{a,b,c}$}
\author{Sandu Popescu$^d$}
\author{Daniel Rohrlich$^e$}

\affiliation{$^a$School of Physics and Astronomy, Tel Aviv University, Tel Aviv 69978, Israel.}
\affiliation{$^b$Institute for Quantum Studies, Chapman University, Orange, California 92866, USA}
\affiliation{$^c$Department of Physics, Schmid College of Science and Technology, Chapman University, Orange, California 92866, USA}
\affiliation{$^d$H. H. Wills Physics Laboratory, University of
Bristol, Tyndall Avenue, Bristol BS8 1TL, UK}
\affiliation{$^e$Physics Department, Ben-Gurion University of the Negev, Beersheba 8410501, Israel.}

\date{Oct 2023}

\begin{abstract}
In a recent paper, PNAS, 118, e1921529118 (2021), it was argued that while the standard definition of conservation laws in quantum mechanics, which is of a statistical character, is perfectly valid, it misses essential features of nature and it can and must be revisited to address the issue of conservation/non-conservation in individual cases. Specifically, in the above paper an experiment was presented in which it can be proven that in some individual cases energy is not conserved, despite being conserved statistically.  It was felt however that this is worrisome, and that something must be wrong if there are individual instances in which conservation doesn't hold, even though this is not required by the standard conservation law. Here we revisit that experiment and show that although its results are correct, there is a way to circumvent them and ensure individual case conservation in that situation. The solution is however quite unusual, challenging one of the basic assumptions of quantum mechanics, namely that any quantum state can be prepared, and it involves a time-holistic, double non-conservation effect. Our results bring new light on the role of the preparation stage of the initial state of a particle and on the interplay of conservation laws and frames of reference. We also conjecture that when such a full analysis of any conservation experiment is performed, conservation is obeyed in every individual case.
\end{abstract}

\maketitle

\newcommand{\tr}{\mbox{Tr}}
\newcommand{\id}{\mbox{\bf I}}
\newcommand{\I}{\mbox{\bf i}}
\newcommand{\ket}[1]{\left | #1 \right \rangle}
\newcommand{\bra}[1]{\left \langle #1 \right |}
\newcommand{\braket}[2]{\left \langle #1 | #2 \right \rangle}
\newcommand{\up}{\uparrow}
\newcommand{\down}{\downarrow}
\newcommand{\beqa}{\begin{eqnarray}}
\newcommand{\eeqa}{\end{eqnarray}}
\newcommand{\beq}{\begin{equation}}
\newcommand{\eeq}{\end{equation}}
\newcommand{\ra}{\rangle}
\newcommand{\la}{\langle}
\newcommand{\vphi}{{\varphi}}
\newcommand{\ttt}{{{\theta_P}}}

\newcommand{\hL}{{\hat L}}
\newcommand{\hA}{{\hat A}}
\newcommand{\hU}{{\hat U}}
\newcommand{\hPi}{{\hat \Pi}}
\newcommand{\hH}{{\hat H}}

Conservation laws are some of the most important laws of physics. Having their origin in the symmetries of nature, conservation laws are present in all our physical theories, from classical mechanics, to relativistic and quantum physics. There are, however, significant differences between what conservation laws mean in these various theories. In quantum mechanics, which is a theory that is non-deterministic at a fundamental level, conservation laws have a different manifestation than in classical mechanics. Indeed, in classical mechanics, in each run of an experiment we can determine the initial value of the conserved quantity and at the end of the experiment we can check that its final value is identical to it. On the other hand, in quantum mechanics where the initial state of a system may not have a well-defined value for the conserved quantity, but some superposition of it, and the outcome of the final measurement cannot, in general,  be predicted from the knowledge of the initial state, one cannot directly apply the same concept of ``conservation" as in classical physics. In face of these difficulties, the concept of conservation that applies to classical mechanics was generalised: the standard definition of conservation used  at present in quantum mechanics is statistical. 

\bigskip
\noindent
However, in a recent paper, \cite{conservation1}, based on ideas first formulated in \cite{conservation2}, we have argued that while the statistical definition is perfectly valid as far as it goes - and it is an extremely useful concept -  it misses essential features of nature and has to be revisited and extended. Reference \cite{conservation1} however didn't go further than pointing out the need for this revision and did not offer a ``solution".  Here we show that taking this idea seriously leads to uncovering some extremely surprising phenomena and challenges one of the basic assumptions of quantum mechanics, namely that any quantum state can be prepared.

\bigskip
\noindent
The argument in \cite{conservation1} stems from the discovery of a particular situation that challenges the standard view.  In particular, we have described there a situation in which a box contains a single particle prepared in a state $|\Psi\ra$ which is a superposition of energy eigenstates, with all energies less than a maximal value $E_{max}$. In some runs of the experiment the particle emerges from the box. When it does this,  it emerges with an energy much higher than $E_{max}$, the maximal energy component present in its initial state, all this while the mechanical system used for extracting the particle - the only other system in the problem - doesn't change its state to account for this energy change. Hence, in the individual cases when the particle emerges from the box, energy is not to be conserved. Similar examples can also be constructed for other conserved quantities such as momentum and angular momentum. So what does this example tell us? 

\bigskip
\noindent
The above example, surprising as it is, is {\it not} in contradiction with the standard definition of conservation laws in quantum mechanics. Quantum conservation laws are statistical, and they refer to many (theoretically infinite) repeated experiments and to the probabilities of the outcomes of these measurements. More precisely, according to the standard definition of  conservation laws in quantum mechanics, a  quantity is said to be conserved if the following is the case: if we prepare an ensemble of systems in the same initial state and immediately after the preparation we measure the quantity of interest, we obtain the same probability distribution of the outcomes as when we first let the systems evolve and we measure this quantity only after the end of the experiment.  What is essential in this definition is that it refers to the {\it entire} ensemble. In our example, if we were to count all runs of the experiment, both those in which the particle emerges out of the box and those in which it didn't, we find that the standard conservation law holds. (After all, it is impossible to be otherwise, by the very way in which conservation laws are defined:  When a quantity commutes with the Hamiltonian - the only situation when the dynamics is such that that quantity should be conserved - the standard conservation law always holds true.)  

\bigskip
\noindent
The problem is that the standard conservation law tells nothing about individual cases. Imagine however that the box contained a particle of energy of order 1 eV (more precisely a particle in a superposition of various energies but absolutely none of them larger than 1 eV) and that in one run of the experiment when we open the box the particle emerges with energy of order of millions of GeV while the mechanism that extracted the particle from the box, the only other system in the problem, didn't change its state. It seems legitimate to worry of what's going on.  To us it seems that something must be wrong if there are individual instances in which conservation doesn't hold, even if over the entire statistical ensemble of cases conservation holds. 

\bigskip
\noindent
Here we revisit the experiment of  \cite{conservation1} and present the solution to that problem. The solution involves two elements: First we avoid the individual-case non-conservation problem presented in \cite{conservation1} by arguing that its setting is unphysical, challenging therefore one of the basic assumptions of quantum mechanics, namely that any quantum state can be prepared. Second, we show that modifying the experiment to make it physical implies that actually  there is conservation in the individual case we consider. This happens via a time-holistic, double non-conservation effect. Our results bring a new light on the role of the preparation stage of the initial state of a particle, as well as on the interplay of conservation laws and frames of reference. We also conjecture that when such a full analysis of any conservation experiment is performed {\it conservation is obeyed in any individual case, not only statistically}. 

\bigskip
\noindent
The paper is organised as follows. We start (the first four sections) by recalling the arguments of \cite{conservation1} for the need to revisit the issue of conservation laws.  We then set the stage (by presenting the basic effect discussed in  \cite{conservation1} but reformulated for the case of angular momentum conservation, which is much simpler as it avoids the complications of time evolution. While the case of angular momentum conservation has been discussed briefly also in an appendix \cite{conservation1} here we give more details which are essential for understanding the new results. The next three sections introduce our basic new conceptual element and present the main result of the paper. The rest of the paper is devoted to discussions and conclusions.

\section*{Superoscillations}

\bigskip
\noindent It is useful to start by recalling the basic experiment in more detail. 

\bigskip
\noindent 
At the core of our examples is the function $f(x)$ 

\beq f(x)=\left( {{1+\alpha}\over 2} e^{i{{x}/{N}}}+{{1-\alpha}\over 2} e^{-i{{x}/ {N}}}\right)^N,~~~\label{basic function}\eeq
with $N$ an integer that we will take as large as we want and with $\alpha$ a real constant with the crucial property that $\alpha>1$. 

\bigskip
\noindent 
This function has the property of ``superoscillations", see \cite{conservation1, conservation2} as well as   \cite{superoscillations1, superoscillations2, superoscillations3} and the extensive review  \cite{superoscillations4}.  It is a widespread belief that a function cannot have features smaller than the smallest wavelength of its Fourier components. Yet this is not true, as shown in  \cite{conservation1, conservation2}.  Superoscillatory functions can oscillate on arbitrary long intervals with much shorter wavelengths than any of their Fourier components. Or, in frequency terms, they can oscillate with much higher frequency than the highest Fourier component. In particular, the function $f(x)$ in (\ref{basic function}),
is a superposition of frequencies strictly limited to the interval $[-1, 1]$. However,
for large $N$, in the region $|x|$ of order of $N^{{1\over2}-\epsilon}$ where $\epsilon$ is an arbitrary small fixed positive constant, the function is 
\beq f(x)\approx e^{i\alpha x}\label{superoscillations}\eeq that is, it oscillates with frequency $\alpha>1$,  faster than the maximal Fourier component.

\bigskip
\noindent To prove the above claims, first, by expanding the binomial, we see that $f(x)$ is a superposition of Fourier components of frequencies ${{2n-N}\over N}$, with $n$ integer, $0\leq  n\leq N$, which are limited to the interval $[-1, 1]$:

\beq  f(x)=\sum_{n=0}^N C_n e^{i{{2n-N}\over N}x}\eeq
with $C_n$ constants
\beq C_n= \left(\begin{matrix} N\\ n\end{matrix}\right)\Big({{1+\alpha}\over 2}\Big)^n\Big({{1-\alpha}\over 2}\Big)^{N-n}.\label{constants}\eeq

However, consider now this function in the region $|x|\leq L$ where $L$ is some fixed number, and let $N$ increase.  (This is a smaller region than $N^{{1\over2}-\epsilon}$ where the superoscillations hold but here the proof is immediate. For the larger region of order $N^{{1\over2}-\epsilon}$ see  \cite{conservation1} as well as Supporting information I.) For large $N$, up to corrections of order ${\cal O}(1/N)$, we can approximate the exponentials by their first order Taylor expansion and obtain 
\beqa &&f(x)\approx\left( {{1+\alpha}\over 2} (1+i{{x}\over N})+{{1-\alpha}\over 2} (1-i{{x}\over N})\right)^N\nonumber\\ & \quad &=\left(1+{{i\alpha x}\over N}\right)^N
\approx e^{i\alpha x}\eeqa
which means that in this region the function oscillates with frequency $\alpha$. 

\bigskip
\noindent Finally note that since the size of the superoscillatory region increases with $N$, the number of superoscillatory wavelengths $\lambda = {{2 \pi}\over {\alpha}}$ in the superoscillatory region can be made as large as we want by taking $N$ large enough.

\section*{The experiment}

\bigskip
\noindent In the present paper we will focus on angular momentum conservation, since it is somewhat simpler than the energy non-conservation example of \cite{conservation1}, though it is constructed along very similar lines.

\bigskip
\noindent 
For simplicity we will also consider the mass of the particle of interest to be large enough and the time scales short enough so that the free Hamiltonian of the particle can be neglected.

\bigskip
\noindent We now construct our basic experiment using a version of the superoscillatory function $f$. Consider a particle moving on a circle and let $\theta$, $-\pi\leq \theta\leq \pi$, denote its position.
\bigskip
\noindent Let the wavefunction be $|\Psi\ra$ which in the angle representation is 

\beq \Psi(\theta)={\cal N}\left( {{1+a}\over 2} e^{i\theta}+{{1-a}\over 2} e^{-i\theta}\right)^N,~~~\label{state_circle}\eeq
where $a$ is an integer larger than 1 and ${\cal N}$ is a normalisation factor.  (The restriction of $a$ to integers is made for mathematical simplicity; the conclusions will not change if we take $a$ to be any rational or real number larger than 1.)

\bigskip
\noindent Consider now the region of small angles $|\theta|$ of order $1/N^{1/2+\epsilon}$.  As we increase $N$, in this region the state $\Psi(\theta)$ presents superoscillations. The proof is immediate:

\bigskip
\noindent Since we will only be interested in angular momentum, once the angular dependence of the wavefunction is given the value of the radius of the circle is irrelevant. However, it is convenient (also with an eye to possible experiments) to think of a circle of radius $R_N=Nr_0 $ with $r_0$ a fixed length. The region of interest, $|\theta|=O(1/N^{1/2+\epsilon})$, corresponds to an arc length $L$ along the circle of order $N^{1/2-\epsilon}r_0$.  In the variable $x=\theta R_N$ that denotes the position along the circle, the situation maps immediately to that investigated in (\ref{basic function})-(\ref{superoscillations}); we can re-express it afterwards in term of angles.

\bigskip
\noindent
Following the same calculation as in (\ref{basic function})-(\ref{superoscillations}), we find that $\Psi(\theta)$ is a superposition of angular momentum eigenstates $e^{im\theta}$ of the angular momentum operator $\hL$ of eigenvalues $m$, with $-N\leq m\leq N$ (where we take $\hbar=1$):

\begin{eqnarray}\Psi(\theta)={\cal N}\sum_{n=0}^N C_ne^{i (2n-N)\theta}={\cal N}\sum_{m=-N}^N c_m e^{i m\theta}\label{sumoverfrequencies_circle}\end{eqnarray}
where $C_n$ is given by eq. (\ref{constants}) with $\alpha=a$ and where $c_m=C_{2n-N}$.

On the other hand  for $|\theta|$ of order $1/N^{1/2+\epsilon}$, in the limit of large $N$, 
\beq \Psi(\theta)\approx{\cal N}e^{i aN\theta}\label{superoscillations_circle}\eeq
a much higher angular frequency, mimicking in this region a particle with angular momentum $aN $, $a$ times higher than the highest angular momentum in the superposition. Note also that  if we take $N$ large enough, we can have the superoscillatory region to extend for as many superoscillatory wavelengths as we want.

\bigskip
\noindent The experiment consists in determining if the particle is in the superoscillatory region. In technical terms we want to measure the projector on that region. When we find the particle there, it has angular momentum approximately  $aN$, which is much higher than all the angular momentum components that it initially had. The question is, where did it take the supplementary angular momentum from? 

\section*{The paradox}

\bigskip
\noindent It is tempting to think that the answer is trivial and straightforward: There are two systems in the problem - the particle and the measuring device used for measuring the position. They interact when the measurement takes place. As the particle increased its angular moment, surely it must have received a kick from the measuring device. In return, the measuring device must have decreased its own angular momentum. Hence, this seems nothing else than a trivial example of conservation of the total angular momentum.  Only that it is not so: {\it  The measuring device could not have provided the necessary angular momentum. In fact, in the limit of large $N$, the measuring device didn't change its state at all.} 

\bigskip
\noindent The argument is simple. Let us compare what happens in two cases: (i) when measurements are made on the particle prepared in 
the superoscillatory state $|\Psi\ra$ which mimics the high angular momentum $aN$ in the region of interest and (ii) when measurements are performed on a particle prepared in an angular momentum eigenstate of eigenvalue $aN$, i.e. when the high angular momentum is genuine. The interaction between the measuring device and the particle takes place in the region of superoscillations. In this region the particle's wavefunction $|\Psi\ra$ looks (up to perturbations that we can make as small as we want by choosing $N$ large enough) like the eigenstate of angular momentum with value $aN$. Suppose now that we perform a measurement of the projection operator on the superoscillatory region. Since the difference between the wavefunction $\Psi$ and the true eigenstate of eigenvalue $aN$ is only encoded in the wavefunction outside the superoscilatory region and since we took the free hamiltonian of the particle to be zero which means that this information cannot propagate, this information it is unavailable to the measuring device and cannot affect the result. Hence the measuring device will behave in the same way in both cases.

\bigskip
\noindent
Let us compare these two cases. In the latter case, the wavefunction is a monochromatic wave of wavelength $\lambda=2\pi R_N/aN=2\pi r_0/a$ (corresponding to the angular momentum $aN$). All that happens when we find the particle in our region of interest, $|x|$ of order $N^{1/2-\epsilon}r_0$, is that the measuring devices simply cuts out of this monochromatic wave a wave-train of length equal to the lengths of this region. Since for increasing $N$ this region contains more and more wavelengths, this wave-train approximates better and better the original monochromatic wave. In other words, if we find the particle in the region of interest, the final wavefunction has angular momentum closer and closer to the original high angular momentum and hence the measuring device doesn't give it any supplementary angular momentum at all.  But then the measuring device couldn't give angular momentum in the first  case either, since in both cases it must behave in the same way.  This is the paradox.

\section*{Modelling the measurement}

\bigskip
\noindent To model the measurement of the projection operator onto the region of interest in such a way as to enable us to see the exchange of angular momentum between the system and the measuring device we proceed as follows.  We take the measuring device to be a second particle which also moves on the circle and which has an internal degree of freedom of this particle, the ``pointer". The location of the measuring device on the circle determines the location of the region where we would like to perform the projection. Let its coordinate be denoted by $\theta_M$, and let its initial wave function be $\phi(\theta_M)$. Let the pointer consist of an internal degree of freedom a described by a 2-dimensional Hilbert space. Initially we take the pointer in the state $|\up\ra$. The initial state of the particle and measuring device is therefore

\beq \Psi(\theta)\phi(\theta_M)|\up\ra\label{total_initial_state_circle}.\eeq

\bigskip
\noindent
What we want to arrange is an interaction that (a) conserves the total angular momentum and (b) the pointer flips to state $|\down\ra$ if and only if the particle is in the superoscillatory region. We can do this by using the interaction Hamiltonian 
\beq H= {{\pi}\over 2}\delta(t)g(\theta-\theta_M;\varphi)\sigma_x\label{hamiltonian_circle}\eeq
where $g(\theta;\varphi)$ denotes a square pulse of width $\varphi$, with $0<\varphi<\pi/2$, i.e. 

\beq g(\theta;\varphi)=\begin{cases} 
1~~ {\rm for}~~ \theta{\rm mod}2\pi \leq \varphi \cr 
0 ~~{\rm for}~~ \theta{\rm mod} 2\pi \geq \varphi
\end{cases}\eeq 
and where the $\sigma_x$ operator flips $\up$ to $\down$ and vice-versa.  

\bigskip
\noindent  Note first that the angular dependence of the Hamiltonian is only via $(\theta -\theta_M){\rm mod}2\pi$, the relative angle between the particle and the measuring device.  This ensures conservation of the total angular momentum. Second, the Hamiltonian is zero for $(\theta-\theta_M){\rm mod}2\pi\geq \varphi$ and proportional to $\sigma_x$ for $(\theta-\theta_M){\rm mod}2\pi\leq \varphi$.  In other words it flips the pointer only when the particle is within a distance $\varphi$ from $\theta_M$, the location of the measuring device, and leaves the pointer unchanged when the particle is further away.  
All that is left is to take $\varphi$ and $\theta_M$ such as to ensure that this region is within the superoscillatory region and captures as much as possible of this region. To do this, we take $\varphi$ of the order of the superoscillatory region and prepare the measuring device in a wavepacket $\phi(\theta_M)$ centred on $\theta_M=0$ and with an appropriately small spread  $\Delta$. For example, we take (up to normalisation) $\phi(\theta_M)=g(\theta_M;\Delta)$ with $\Delta<<\varphi$, so that the interval $|\theta|\leq|\varphi+\Delta|$ is still in the superoscillatory region.

\bigskip
\noindent  
The corresponding unitary time evolution operator over an infinitesimal time $\tau$ is
\beqa &&\hU(\tau, 0)=e^{i\int_0^{\tau} \hH(t)dt}=e^{i{\pi\over 2}g(\theta-\theta_M; \varphi)\sigma_x}\nonumber\\&&=\cos\Big({\pi\over 2}g(\theta-\theta_M; \varphi)\Big)+i\sigma_x\sin\Big({\pi\over 2}g(\theta-\theta_M; \varphi)\Big)\nonumber\\
&&=1-g(\theta-\theta_M; \varphi)+ig(\theta-\theta_M; \varphi)\sigma_x,\eeqa
where in the last equality we used in the trigonometric functions the definition of the function $g$.

\bigskip
\noindent  
For any arbitrary initial state $\psi(\theta)$ of  the particle, the state of the particle and the measuring device is a superposition of two terms, one with the pointer flipped (for $(\theta-\theta_M){\rm mod}2\pi\leq \varphi$) and one with the pointer not flipped:
\beqa&&\psi(\theta)\phi(\theta_M)|\up\ra \rightarrow \hU\psi(\theta)\phi(\theta_M)|\up\ra=\nonumber\\&&\Big(1-g(\theta-\theta_M; \varphi)\Big)\psi(\theta)\phi(\theta_M)|\up\ra+\nonumber\\&&ig(\theta-\theta_M; \varphi)\psi(\theta)\phi(\theta_M)|\down\ra.\eeqa
As expected, the term with the flipped pointer is (up to normalisation) simply the arbitrary original state $\psi$ truncated to a region of size $\varphi$ centred on $\theta_M$, that is, the projection of the original state onto this region.  

\bigskip
\noindent  Let's now take the arbitrary state $\psi$ to be the superoscillatory state $\Psi$. In this case, when the pointer has flipped, the state of the particle and measuring device is (up to normalisation)

\beq ig(\theta-\theta_M; \varphi)\Psi(\theta)\phi(\theta_M)=ig(\theta-\theta_M; \varphi)\Big(e^{iaN\theta}+O({1\over N})\Big)\phi(\theta_M)\label{projected state}\eeq
using the approximation of $\Psi$ for the small angles and the fact that for larger angles $g(\theta-\theta_M; \varphi)=0$.

\bigskip
\noindent  The crucial thing, as we explained above in the previous section, is that when the particle was found in the superoscillatory region, the particle-measuring device state is, up to approximations of $1/N$ which can be made as small as we want, the same as when the initial state of the particle were  the genuine high angular momentum eigenstate $e^{iaN\theta}$. So whatever happens to the measuring device when the initial state is the bona-fide high angular momentum state, happens also when the initial state is our ``fake" high angular momentum state.

\bigskip
\noindent  Let us now study the particle-measuring device angular momentum exchange. Consider first what would happen if the particle were to start in the high angular momentum state $aN$ and the measuring device pointer is flipped, meaning that the particle is found in the $|\theta|$ region where our function $\Psi$ superoscillates.  To see the angular momentum exchange in this situation we decompose the interaction term $g(\theta-\theta_M; \varphi)$ in its Fourier transform and obtain

\beqa&& g(\theta-\theta_M; \varphi)e^{iaN\theta}\phi(\theta_M)\nonumber\\&&= \sum_{k=-\infty}^{\infty} {\rm sinc}(k\varphi) e^{ik(\theta-\theta_M)}e^{iaN\theta}\phi(\theta_M) \nonumber\\ &&=\sum_{k=-\infty}^{\infty} {\rm sinc}(k\varphi) e^{i(aN+k)\theta}e^{-ik\theta_M}\phi(\theta_M) \label{angular momentum conservation shifts}\eeqa
where we used 
\beq g(\theta;\varphi)=\sum_{k=-\infty}^{\infty} {\rm sinc}(k\varphi) e^{ik\theta}.\eeq 
Here $e^{i(aN+k)\theta}$ is an eigenstate of the angular momentum of the particle corresponding to the eigenvalue $aN+k$  and $e^{-ik\theta_M}\phi(\theta_M)$ if a state that is identical to $|\Phi\ra_M$, the initial state of the measuring device, but shifted down in angular momentum by $k$.

\bigskip
\noindent  In the Dirac notation that will be more convenient in the next sections, for the cases when the particle is found in the superoscillatory region (i.e. the measuring device pointer is flipped), the measurement time evolution is given by

\beq |aN\ra_p|\Phi\ra_M\rightarrow \sum_{k=-\infty}^{\infty} {\rm sinc}(k\varphi)|aN+k\ra_p|\Phi -k\ra_M\label{angular momentum conservation shifts dirac}\eeq
where by $|\Phi -k\ra_M$ we denote a state that is identical to $|\Phi\ra_M$, the initial state of the measuring device, but shifted down in angular momentum by $k$.

\bigskip
\noindent   What eq.(\ref{angular momentum conservation shifts}) and (\ref{angular momentum conservation shifts dirac}) show is that when the particle emerges with the same high angular momentum $aN$ as it started with (that is, corresponding to $k=0$) the measuring device doesn't change its angular momentum at all. This is very much expected: since the particle didn't change its angular momentum, there is no reason for the measuring device to provide it any supplementary angular momentum, so no change of the angular momentum of the measuring device should occur.  When the particle emerges with other angular momentum value $aN+k$ with $k\neq0$, which occurs due to the truncation of the initial monochromatic wavefunction, the measuring device changes its angular momentum by $-k$ to compensate for the change. In other words, nothing changes in the measuring device when the particle emerges with precisely the high angular momentum $aN$; there are only changes to account for the deviations of the particle's angular momentum from this value. Incidentally, we also note that these truncation disturbances become relatively less and less important with increasing $N$, since while the central angular momentum value increase as $aN$, the only significant deviations from it are only of order $|k|=O(N^{1/2-\epsilon})$. 

\bigskip
\noindent 
Now, had the particle started in our initial state $\Psi$, the measuring device would have reacted in the same way, up to corrections of $1/N$ as described in (\ref{projected state}), so when found in the superoscillatory region we have
\beq |\Psi\ra_p|\Phi\ra_M\rightarrow \sum_{k=-\infty}^{\infty} {\rm sinc}(k\varphi)|aN+k\ra_p|\Phi -k\ra_M +O({1\over N}).\eeq

This means, in particular, that up to $1/N$ corrections, when the particle emerges with the high angular momentum value $aN$ (with $a>1$), the measuring device doesn't provide any angular momentum to the particle, despite the fact that maximal initial angular momentum of the particle is only $N$. Where does the difference $aN-N$ come from in this case? This is our paradox. 

\bigskip
\noindent 
The following sections give the solution.

\section*{Preparation}

\bigskip
\noindent
Let us now discuss a seemingly unrelated issue: How is this special state of the particle prepared in the first place? The reason the preparation is not completely trivial is that we are dealing with a conserved quantity, and that imposes some constraints. 

\bigskip
\noindent
In our original paper \cite{conservation1} we started by discussing a box containing a photon prepared in a similar special superposition of energy eigenstates. Presumably, a laser should have been employed to deliver the photon into the box. But the total energy is a conserved quantity, hence when the laser emits the photon with a certain energy, the laser must lose the same amount. (N.B.  Actually, in general the laser is connected to other objects - optical table, external power supply. It is the energy of this whole system that changes to compensate the change in the energy of the photon. For simplicity, here we call ``laser" this entire system.) 

\bigskip
\noindent
As the photon is in a superposition of energy eigenstates, the laser will get entangled with it, each energy eigenstate of the photon being correlated to another state of the laser, shifted down in energy by the energy lost to the photon. The state of the photon is thus not a pure state, as we desired. However, if the state of the laser has a wide and smooth enough distribution of energy, a shift in energy doesn't modify the state considerably and the entanglement is minimal. This is how non-monochromatic but almost pure states of light are routinely prepared. In our present case, we could similarly imagine a ``preparing device", moving on the same circle as the particle, and isolated from everything else, that releases the particle on the circle that we consider. Or, for greater simplicity, imagine that we start with the particle in the initial state of zero angular momentum $|0\ra_p$ and the preparer boosts its angular momentum appropriately. Angular momentum being a conserved quantity, the state of the preparer also changes.  Again, the particle and the preparing device get entangled. Nevertheless, by choosing the the initial state of the preparation device appropriately we can make this entanglement arbitrarily small.

\bigskip
\noindent In other words, instead of preparing the pure state 
\beq |\Psi\ra_p=\sum_{m=-N}^N c_m|m\ra_p\eeq 
what the actual preparation procedure does is to take the initial particle-preparer state $ |0\ra_p|\Phi\ra_P$ to the final, entangled state $|\chi\ra_{p, P}$, via the angular momentum conserving transformation

\beq  |0\ra_p|\Phi\ra_P\rightarrow |\chi\ra_{p, P}=\sum_{m=-N}^N c_m|m\ra_p|\Phi -m\ra_{P}\label{entangled preparation}\eeq
where $|\Phi -m\ra_{P}$ is the initial state of the preparer shifted down in angular momentum by $m$. 

As discussed above, to make the reduced density matrix of the photon in the two-party entangled state (\ref{entangled preparation}) approximate that corresponding to the pure state $\Psi$ we need to chose the initial state of the preparer, $|\Phi \ra_{P}$, as a wide and smooth enough superposition of angular momentum so that for all $m$, with $-N\leq m\leq N$ the shifted states
 $|\Phi -m\ra_{P}$ are almost identical to the initial state $|\Phi\ra_{P}$ 
 (i.e. $_{P}\la\Phi-m|\Phi\ra_{P}\approx 1$).  To achieve this goal, we take $|\Phi\ra_{P}$  to be narrow in its angular distribution. Indeed, the angle $\ttt$ that denotes the location of the preparer is conjugated to its angular momentum, so a narrow distribution of $\ttt$ ensures a wide and smooth distribution of angular momentum.

\bigskip
\noindent Going beyond the above purely mathematical argument of why we want to take $\Phi_{P}(\ttt)$  with a narrow distribution of $\ttt$, here are two physical reasons. 

\bigskip
\noindent First, since total angular momentum is conserved, the Hamiltonian cannot depend on the absolute value of the angle of the particle or of the preparing device but only on their relative angle $\theta-\ttt$. Hence, for each value of $\ttt$ corresponds a wavefunction of the particle appropriately shifted in space.  This, of course, means entanglement between the particle and the preparing device, which we want to avoid. To limit this entanglement we need to reduce the uncertainty in $\ttt$, hence to take for the state of the preparer $\Phi(\ttt)$ a wavepacket narrow in angular distribution. 

\bigskip
\noindent Another way to look at the problem is to note that we want to place the state of the particle at a precise position on the circle - if there is uncertainty in its positioning, as if it would be the case if $\ttt$ were to have a large uncertainty, it will wash out the finer details of its space dependence, so we will not be able to see the short wavelength oscillations we are interested in. To ensure this, we need, again, to take for $\Phi(\ttt)$ a wavepacket narrow in angular distribution. 

\bigskip
\noindent For an explicit example let $\Phi(\ttt) $ be a top-hat function with spread $2\eta$, i.e. 
\beq \Phi(\ttt)={1\over{\sqrt{2\eta}}}g(\ttt;\eta).\eeq Then the scalar product 
\beqa&&_{P}\la\Phi-m|\Phi-m'\ra_{P}=\int_{-\pi}^{\pi}e^{im\ttt}\Phi^*(\ttt) e^{-im'\ttt}\Phi(\ttt) d\ttt\nonumber\\&&={1\over{2\eta}}\int_{-\eta}^{\eta}e^{im\ttt}e^{-im'\ttt}d\ttt={{\sin\Big((m-m')\eta\Big)}\over{(m-m')\eta}}\label{scalar products}\eeqa
where we have used the angular representation $e^{-im'\ttt}\Phi(\ttt)$ of $|\Phi-m'\ra_{P}$.

\bigskip
\noindent 
Since in our case the angular momenta of interest are limited to the interval $-N\leq m, m'\leq N$, the difference is bounded, $|m-m'|\leq 2N$. Hence by choosing  $\eta$ small enough, we can make all the scalar products (\ref{scalar products}) as close to 1 as we wish and hence, from (\ref{entangled preparation}), the state of the particle as close to $\Psi$ as we wish.

\section*{The final state of the preparer}

\bigskip
\noindent We have now arrived to the crucial point of our paper. Although in our preparation stage we have not prepared the pure state $\Psi$, the state that we prepared has all the ingredients essential for our purpose:

\begin{itemize}
\item The particle's state has only angular momentum components in the interval $[-N, N]$. Note that although the particle's state is only an approximation of the pure state $\Psi$, the condition that there are no angular momentum components larger than $N$ in its decompositions is strictly valid, as in $\Psi$.
\item By choosing $\eta$ appropriately small we can make the particle's state approximate as well as we want the pure state $\Psi$, In particular this new state of the particle presents the same superocillations as $\Psi$: in the region $|\theta|=O(1/N^{1/2+\epsilon})$ it looks approximately like $e^{iaN\theta}$, corresponding to angular momentum $aN$, with $a>1$, larger that all angular momentum components in its decomposition.
\end{itemize}

\bigskip
\noindent
Due to the above, in the subsequent stage of the experiment - the measurement stage -  all will be as close as we want to what happens when starting with $\Psi$. That is, when the particle is found in the superoscillatory region, its angular momentum is close to $aN$, much larger than its maximal initial value of $N$. However the measuring device doesn't provide the angular momentum difference to compensate for the increase of the particle's angular momentum from its maximal initial value of $N$ to around $aN$. More precisely, when the particle is found with angular momentum $aN$ the state of the measuring device doesn't change at all, so it doesn't give the particle any angular momentum whatsoever.  When the particle is found with angular momentum different from $aN$ (which occurs with non-negligible probability only  for deviations of order $N^{1/2}$), the measuring device provides angular momentum  to compensate for this difference, but not for the increase from the maximal initial value of $N$ to $aN$. Hence our paradoxical effect stands: the particle can emerge with an angular momentum much larger than any angular momentum in its initial state without the measuring device - the only system with which it interacted - to give it the difference. 

\bigskip
\noindent Let's now see what happens to the preparer. 

\bigskip
\noindent  At first sight, as far as angular momentum is concerned, the preparer's role is rather trivial: it provides the angular momentum gained by the particle from its initial zero angular momentum state, $|0\ra_p$ to the desired superposition. Indeed, that was the whole point of the previous section. As (\ref{entangled preparation}) shows, if after the preparation we measure the angular momentum of the particle and find it to be $m$, meaning that the particle gained angular momentum $m$, then the preparer is in the state $|\Phi-m\ra_{P}$, whose angular momentum probability distribution is identical to that in the initial state $|\Phi\ra_{P}$, but shifted by $-m$. This is a trivial case of angular momentum conservation, as discussed above.

\bigskip
\noindent 
Importantly for us here, since the particle's state only has components  limited to the interval $[-N, N]$, the preparer never provides more than angular momentum $N$ (i.e. the maximal down shift of the angular momentum distribution of the preparer is $-N$).

\bigskip
\noindent 
Our experiment, however, requires that after preparation we first check if the particle is in the superoscillatory region and only then we measure its angular momentum.  So suppose we find the particle in a large spatial interval in superoscillatory region and that the subsequent measurement of the particle's angular momentum yields the value $k$.  The preparer's state in this situation is (up to normalisation)
\beq _p\la k|\hPi |\chi\ra_{p, P}\label{preparer state superoscillation}\eeq
where $\hPi$ is the particle projection operator onto that interval.

To evaluate the preparer's state it is convenient to express $ |\chi\ra_{p, P}$ as
\beqa& &|\chi\ra_{p, P}=\sum_{m=-N}^N c_m|m\ra_p|\Phi -m\ra_{P}\nonumber\\&=&\sum_{m=-N}^N c_me^{im{\hat\theta}_P} |m\ra_p|\Phi\ra_{P}\nonumber\\&=& e^{i\hL{\hat{\theta}_P}} \sum_{m=-N}^N c_m|m\ra_p|\Phi\ra_{P}=e^{i\hL{\hat{\theta}_P}} |\Psi\ra_p|\Phi\ra_{P}\label{entangled preparation alternative}\eeqa
where $\hL$ is the particle angular momentum operator and ${\hat{\theta}_P}$ the particle angle operator.

Using (\ref{entangled preparation alternative}) and the crucial fact that $|\Phi\ra_{P}$, the initial state of the preparer, has only a very narrow angular distribution around $\ttt=0$ we obtain
\beqa &&_p\la k|\hPi |\chi\ra_{p, P}=~_p\la k|\hPi e^{i\hL{\hat{\theta}_P}} |\Psi\ra_p|\Phi\ra_{P}\nonumber\\&\approx&~_p\la k|\hPi (1+i\hL{\hat{\theta}_P}) |\Psi\ra_p|\Phi\ra_{P}\nonumber\\
&=& _p\la k|\hPi  |\Psi\ra_p\Big(1+i{{ _p\la k|\hPi \hL |\Psi\ra_p}\over{_p\la k|\hPi  |\Psi\ra_p}}{\hat{\theta}_P}  \Big)|\Phi\ra_{P}\nonumber\\
&\approx& _p\la k|\hPi  |\Psi\ra_pe^{i{{ _p\la k|\hPi \hL |\Psi\ra_p}\over{_p\la k|\hPi  |\Psi\ra_p}}{\hat{\theta}_P} }|\Phi\ra_{P}\nonumber\\
&= &_p\la k|\hPi  |\Psi\ra_p|\Phi-{ {_p\la k|\hPi \hL |\Psi\ra_p}\over{_p\la k|\hPi  |\Psi\ra_p}} \ra_{P}\label{final preparer state}
\eeqa
where we used first order approximations in the small angle $\ttt$. 

\bigskip
\noindent The last line of (\ref{final preparer state}) shows that (up to normalisation), the final state of the preparer is equal to its initial state but displaced down in angular momentum by ${ {_p\la k|\hPi \hL |\Psi\ra_p}/{_p\la k|\hPi  |\Psi\ra_p}}$. Let us evaluate this quantity.

\beqa &&{{ _p\la k|\hPi \hL |\Psi\ra_p}\over{_p\la k|\hPi  |\Psi\ra_p}}={{\int_{\rm super} e^{-ik\theta} (-i){{\partial \Psi(\theta)}\over{\partial \theta}}d\theta}\over{ \int_{\rm super} e^{-ik\theta}  \Psi(\theta)}}\nonumber\\&&={{\int_{\rm super} e^{-ik\theta} (-i){{\partial}\over{\partial \theta}}(e^{iaN\theta}+O(1/N))d\theta}\over{ \int_{\rm super} e^{-ik\theta}  (e^{iaN\theta}+O(1/N))d\theta}}\nonumber\\&&=aN+O(1/N).
\eeqa

Hence, when the particle is found in the superoscillatory region and a subsequent measurement of its angular momentum yields $k$, the corresponding state of the preparer is 
\beq |\Phi-aN \ra_P,\label{preparer angular momentum supply}\eeq
up to corrections of order O(1/N ).

\bigskip
\noindent 
This is the key result of our paper. We will discuss its significance in the next sections, but before that let us note a few more details. First, the final state of the preparer is not affected by the precise localisation of the projected space interval in the superoscillation area. (That's why we denoted the integration domain simply as ``super" and not by indicating an exact location). This is important for our further analysis where we consider the projection being made by the interaction with the position measuring device. As discussed before, the projection realised by the position measurement is not on some precise location on the circle, but on an interval centred around $\theta_M$, the location of the measuring device.  Even though we can make $\theta_M$ as precisely localised on the circle as we want, we can never achieve infinite precision, so the projection intervals will differ slightly from one another.  Since it is not sensitive to the precise location of this space interval, our result about the final state of the preparer applies to all these particular projections.

\bigskip
\noindent Second, we note that the final state of the preparer is independent of the particular angular momentum value $k$ with which we find the particle after localising it in the superoscillatory region. In all the cases when the particle is found in the superoscillatory region, the final state of the preparer is the same (\ref{preparer angular momentum supply}).

\section*{A second non-conservation: ``preparation non-conservation''}

\bigskip
\noindent 
Let us now analyse our key result. What eq.(\ref{preparer angular momentum supply}) shows is that  in all of the cases where the particle was later found to be in the superoscillatory region, the final state of the preparer is equal to its initial state but shifted down in angular momentum by $aN$. That is,  in all these cases, the preparer loses $aN$ angular momentum. Yet, the particle started in the state of zero angular momentum, $|0\ra_p$ and the preparation brings it in a state (\ref{entangled preparation})  whose angular momentum components are limited to the interval $[-N,N]$. Thus, the angular momentum lost by the preparer is {\it not} transferred to the particle. We have found therefore another instance of angular momentum non-conservation in our cases of interest. 

\bigskip
\noindent This ``preparation non-conservation" is, if anything, more subtle than the non-conservation that occurs during the subsequent position measurement. The crucial point is that we cannot see this non-conservation right away but we have to {\it wait until after the position measurement}. Indeed, while there are, of  course, many things we can find out immediately after the preparation, without any need to wait, they are not enough for spotting the preparation non-conservation. For example, we can, if we wish, certify that the particle has only angular momentum components in the interval $[-N,N]$. In other words, we certify that at this stage the particle did not receive any large momentum such as $aN$. Importantly, we can certify this {\it without affecting the experiment}. This can be done by simply measuring the projector on this low angular momentum subspace. Since the state of the particle is an eigenstate of this projector, this measurement will not affect our experiment at all. Furthermore, we also can, without waiting, measure the angular momentum that the preparer has after the preparation. Since the preparer finished its interaction with the particle, and never interacts with it again, this measurement doesn't affect our experiment either. What we {\it cannot do right away} after the preparation and have to wait until after the position measurement, is to know whether the particle will be found in the superoscillatory region or not. Not knowing this, we do not know which of the individual runs of the experiment correspond to our case of interest, hence we do not know which results of the measurement of the angular momentum of the preparer should be included. 

\section*{Putting all together: overall conservation}

\bigskip
\noindent Let us now put everything together. Our entire experiment consists of:
\beqa &&{\rm Prepration} \rightarrow {\rm Measurement~of~position}\rightarrow{\rm Selection}\nonumber\\ &&{\rm of~cases~of~particle~in~superoscillatory~region}\rightarrow\nonumber\\&&{\rm Measurement~of~angular~momentum~of~particle}.\nonumber\eeqa
To these we may add, between the preparation and position measurement, a verification that the particle angular momentum is restricted to the interval $[-N,N]$, i.e. measuring the projector on this interval.

\bigskip
\noindent
The corresponding time evolution,  restricting ourselves only to the cases when the particle is found at the end of the experiment in the superoscillatory region, is
\beqa &&|0\ra_p|\Phi\ra_P|\phi\ra_M\rightarrow\nonumber\\&& \rightarrow|\chi\ra_{p, P}|\phi\ra_M=\sum_{m=-N}^N c_m|m\ra_p|\Phi -m\ra_{P}|\phi\ra_M\Rightarrow\nonumber\\\Rightarrow
&&\sum_{k=-\infty}^{\infty}{\rm sinc}(k\varphi)|aN+k\ra_p|\Phi-aN\ra_P|\phi-k\ra_M,\eeqa
where the simple arrow "$\rightarrow$" describes unitary evolution while the double arrow "$\Rightarrow$" describes unitary evolution plus selection and $\varphi$ is the size of the angular interval within the superoscillatory region onto which the position projection measurement is performed.

\bigskip
\noindent
What the above equation shows is that the particle starts with zero angular momentum (i.e. in state $|0\ra_p$) and ends up gaining angular momentum $aN+k$, while the preparer provides $aN$ and the position measuring device provides $k$. Yet, the preparation stage alone doesn't provide the particle with the high angular momentum $aN$: at the end of the preparation stage the particle's maximal angular momentum is only $N$.

\bigskip
\noindent
This is the main result of our study:  

\begin{quote}
The two non-conservation effects - the ``preparation non-conservation'' and the ``measurement non-conservation'' - compensate each other and lead to overall angular momentum conservation {\it even when we restrict us to the sub-ensemble of cases when the particle is found in the superoscillatory region}. That is, conservation holds for these {\it individual cases}, not only statistically. 
\end{quote}

\section*{Discussion I}

\bigskip
\noindent Let us now discuss the above result. First, three simple observations. Given that the particle is found in the superoscillatory region:

(i) It is the preparer that delivers the angular momentum $aN$, allowing the particle to behave {\it at the end of the experiment} as a particle with genuine angular momentum $aN$. 

(ii) When after the position measurement the particle is found with angular momentum $aN$, the measuring device doesn't provide any angular momentum at all. 

(iii) When the particle is found with angular momentum different from $aN$, say $aN+k$, the position measuring device delivers angular momentum $k$. This is precisely what it would have done, had the particle been a genuine high angular momentum state before the position measurement, to account for the disturbance done to that monochromatic wave by truncating it to a finite wave train. 

\bigskip
\noindent More importantly, this is not a simple conservation - it is the composite of two non-conservations. 
It also has a ``holistic" time character: As we noticed before, by doing or not doing the position projection measurement we can change the overall angular momentum transfer from the preparer to the particle, despite that (a) at this stage the particle and the preparer no longer interact and (b) that we can certify that after the interaction with the preparer the particle does not have any high angular momentum at all.  Which leads us to ask: despite the overall conservation, where  from and how precisely did the particle get the high angular momentum, if it didn't have it before interacting with the position measuring device and  the position measuring device didn't change its state? Where has the high angular momentum lost by the preparer been stored until it finally materialised in the particle?

\bigskip
\noindent
Note that both the preparer and the position measuring device play crucial roles: the preparer is the source of the high angular momentum gained by the particle, while the position measuring device gives an ``umbrella" under which this transfer can occur at a time later than the one when the particle and the preparer interacted.

\bigskip
\noindent It is also instructive to contrast the above story with what happens in a related, but fundamentally different, case. Suppose that instead of preparing the particle in our special superposition, i.e. instead of the transformation (\ref{entangled preparation}), we simply use the preparer to prepare a state of well defined angular momentum $m$, with $-N\leq m\leq N$, and then subject it to the same experiment as before. That is, after preparation we measure the projection operator on what previously was the superoscillatory region and, after that, if we found the particle in this region, we measure the angular momentum of the particle. In this region the particle's state oscillates with spatial frequency corresponding to the angular momentum $m$. Yet, after the truncation of the wavefunction, it is still possible to find the particle with angular momentum $aN$, much larger than its initial angular momentum $m$. This happens with small probability, corresponding to the tail of the frequency probability due to truncation, but still, with non-zero probability. (Contrast this with the probability of  finding the particle with momentum close to  $aN$ in the superoscillatory case (given that we found the particle in this region) which is close to 1.)  Where did the particle get its angular momentum from now? 

\beqa &&|0\ra_p|\Phi\ra_P|\phi\ra_M\rightarrow |m\ra_p|\Phi-m\ra_P|\phi\ra_M\Rightarrow\nonumber\\   &&\sum_{k=-\infty}^{\infty} e_k |aN+k\ra_p|\Phi-m\ra_P|\phi-(aN+k-m)\ra_M\nonumber\\\eeqa
with $e_k={\rm sinc}((aN+k-m)\varphi)$.

\bigskip
\noindent
What we see here is a trivial case of angular momentum conservation. The particle gains angular momentum $aN+k$, by a two stage process. In the first stage it gains $m$ and the preparer provides it, by shifting its state to  $|\Phi-m\ra_P$, i.e. down in angular momentum by $m$. No non-conservation now at this stage. In the second stage, the position measuring device disturbs this intermediate particle state and brings its angular momentum from $m$ to $aN+k$, while losing itself $aN+k-m$, hence a trivial conservation again.

\bigskip
\noindent Coming back to our example, one can ask how is it possible that the position measurement and associated selection can affect what has happened to the preparer at an earlier time? Moreover, why does the preparer matter at all? After all, once a state, say $|\Psi\ra$, is prepared, it doesn't matter how it was prepared. The answer is that in our modified experiment we did not prepare a pure state. We have {\it not} prepared the special state $|\Psi\ra$. What we have prepared is an entangled state $|\chi\ra_{p, P}$, (\ref{entangled preparation}) between the particle and the preparer, with the property that the reduced density matrix of the particle can be made as close as we want to the pure state $\Psi$. Nevertheless, the particle is entangled with the preparer. True, $|\chi\ra_{p, P}$ is a very weakly entangled state: each angular momentum component $|m\ra_p$ is correlated with a corresponding preparer state $|\Phi-m\ra_P$ which is identical to the initial state of the measuring device shifted in angular momentum by $m$.  As discussed in the preparation section,  we can make these states of the preparer as close to each other as we want,  $_P\la\Phi-m|\Phi-m'\ra_P\rightarrow 1$, by making them wider and wider spread in angular momentum, so that $|\chi\ra_{p, P}\rightarrow|\Psi\ra_p|\Phi\ra_P$. This means that
the state of the particle can be made as close as we want to the pure state $\Psi$. Hence for its subsequent behavior the particle behaves as close as we want to how it would in the pure state $\Psi$. Yet, for questions related to angular momentum conservation there is an enormous difference between the entangled state $|\chi\ra_{p, P}$, and the pure state $|\Psi\ra_p|\Phi\ra_P$. Indeed, although the states $|\Phi-m\ra_P$ can be made as close to each other as we want, 
their average angular momenta remain finitely different from each other,  $L_0-m$ versus $L_0-m'$ with $L_0$ being the average angular momentum of $|\Phi\ra_P$. This is the reason why this entanglement, while minimal, is however essential. It is essential even for the simple case when we just prepare the initial state $|\chi\ra_{p, P}$ and then we immediately verify that the preparation conserves angular momentum.  It is even more important in our special case. Post-selecting for the particle being in the superoscillatory region and then for a particular final angular momentum produces a superposition of these shifted wavefunctions, which, by interference, lead to a wave function with a much larger shift in angular momentum, as discussed in the previous sections.

\bigskip
\noindent 
Of course, while the entanglement with the preparer explains the mathematical machinery that allows our results to exist, it is by no means, by itself,  an explanation for why this effect happens. It doesn't tell why it is the case that this entanglement is exactly such that angular momentum conservation is valid for the {\it individual cases} when the particle is found in the superoscillatory region, while the only thing demanded by the standard conservation laws is statistical conservation over the entire ensemble. The main principle, as we see it, is that we should demand more of the conservation laws than the statistical conservation. 

\section*{Discussions II. The pure state $\Psi(\theta)$ doesn't exist in nature}

\bigskip
 \noindent 
 One could, of course, try and bring a counterargument to our general line of thinking. One could ask what would happen if the particle were initially exactly in the pure state $\Psi(\theta)$? Then there would be no preparer to provide the required supplementary angular momentum and we would not have angular momentum conservation in our individual case. Would this not invalidate our arguments? 
 
 \bigskip
 \noindent 
Yes, if the particle could start in the initial state $\Psi(\theta)$ there would be genuine non-conservation in the individual cases presented in our original experiment, as described in \cite{conservation1}. It would then follow that conservation doesn't need to hold in individual cases but only statistically, no matter how unpalatable this may seem to us.  But there is a way out. This brings us to the last key element of our paper. We claim that:
 
\begin{quote}The pure state $\Psi(\theta)$ is unphysical and doesn't exist in nature. 
\end{quote}

 \bigskip
 \noindent Arguing that $\Psi(\theta)$ is unphysical and doesn't exist in nature seems to contradict one of the basic postulates of quantum mechanics, namely that any normalised wavefunction is a legitimate state for a quantum particle.  The answer is that the issue is one concerning frames of reference.  The wavefunction $\Psi(\theta)$ has no meaning because the angle $\theta$ has, strictly speaking, no meaning. 
 
 \bigskip
 \noindent It is standard in quantum mechanics to consider wavefunctions of variables $x$, $y$, $z$ or of angle $\theta$ and so on. But it is well understood that what they do represent are not some absolute values but they are coordinates measured relative to some frame of reference, say the walls of the laboratory. The reason why one doesn't always mention this explicitly is that in general this doesn't significantly affect the results and can be ignored. In our case however it matters significantly, as we have shown. The preparer acts as a reference frame and this is why it should always be present. 

 \bigskip
 \noindent 
Lest it appears that the statement that ``we need to always use a frame of reference in quantum mechanics in order to correctly describe conservation laws'' is trivial, we want to emphasise that this is by no means so. Far from it. It is in fact a very subtle issue, which only now it is becoming apparent. Indeed, there is absolutely nothing in standard quantum mechanics that requires us to refer to a frame of reference to ensure that the usual, statistical conservation laws hold.  Indeed, in the standard quantum mechanical formalism nothing prevents us to write an arbitrary wavefunction of two particles, say $\Psi(x_1)\Phi(x_2)$, and as long as they interact via a potential that depends only on their relative distance, $V(x_1-x_2)$, the standard statistical momentum conservation law holds {\it exactly}. Nothing else is needed. There is no need to refer to any other frame of reference or any ``preparation'' device etc. Similarly, no consideration of a reference frame is needed to ensure the statistical conservation of angular momentum or energy, etc, and they apply to all wavefunctions.  We could, if we fancy doing so, refer all the position or time coordinates to some other physical system, but there is no need whatsoever to do it. What we discovered here is therefore a new fundamental aspect of conservation laws:

\begin{quote}If we insist on conservation at an {\bf individual} level, we need to take the issue of frames of reference and of preparation devices into account, while this  is not necessary if we only demand conservation at statistical level.
\end{quote}

\bigskip
 \noindent 
Moreover, and crucially important: the mere thing that one needs a frame of reference, and hence a preparer, is only a {\it necessary} but by no means {\it sufficient} condition to imply conservation at the level of individual cases. This is by no means the ``explanation'' for the effect.  What is required is the entire subtle interplay between the preparer and the particle and their nonlocal-in-time angular momentum transfer via the measuring device which acts as an umbrella that makes this transfer possible by covering otherwise inevitable causality violations. 

\section*{Conclusions: The meaning of quantum conservation laws and the basic postulates of quantum mechanics}

\bigskip
\noindent 
Let us zoom out from the particular example discussed here.  We have argued that while the standard formulation of conservation laws in quantum mechanics, which is of a {\it statistical nature},  is perfectly valid as far as it goes, it is incomplete and needs to be revisited, specifically that we need to go beyond the statistical formulation and consider individual cases as well. In particular, we have argued that if we find  {\it in one particular run} of an experiment,  that, say, when we open a window of a box containing a single particle of energy strictly smaller than 1 eV  and the particle emerges with energy of order of millions of GeV, it is legitimate to ask where did this increase in energy come from {\it in this individual case}. Had we ignored this question on grounds that the standard, statistical, conservation law is all there is, we would have not discovered the strange conservation effect for the individual case presented here, and hence missed a lot of interesting physics. We would  have also missed the implications for the need to explicitly take into account the issue of the frames of reference involved in the preparation of the initial state, issue which does not arise when only the standard statistical conservation law is  concerned but which arises when the conservation in individual cases is concerned, neither the fact that certain quantum states are unphysical, while there would be considered perfectly valid in standard quantum mechanics. We take this as a good indicator of the validity of this line of enquiry, and believe these results are only the tip of an iceberg, part of a more general structure concerning conservation laws. In particular, we conjecture that when such a full analysis of any conservation experiment is performed {\it conservation is obeyed in any individual case}, not only statistically. 

\section*{Acknowledgements}
SP acknowledges the support of the Advanced ERC Grant FLQuant.

\section*{Supplementary Information I}

\bigskip
\noindent
The proof of the superoscillatory nature of $f(x)$ over the range of $|x|\leq O(N^{1/2-\epsilon})$ where $\epsilon$ is an infinitesimally small, positive constant  has been given in \cite{conservation1}. We include it here for convenience. 

\bigskip
\noindent
{\it Theorem:} In the limit of large $N$, for any $|x|\leq {\cal O}(N^{1/2-\epsilon})$ with $\epsilon$ positive
and infinitesimally small, and $\alpha>1$, the function $f(x)$ can be approximated by
\beq f(x)\approx e^{i\alpha x}.\eeq

\bigskip
\noindent
{\it Proof}. Let us express $f$ by using its absolute value and phase: \beqa &&f(x)=\Big( {{1+\alpha}\over 2} e^{i{x\over N}}+ {{1-\alpha}\over 2} e^{-i{x\over N}}\Big)^N=\nonumber\\&&\Big(\cos{x\over N}+i\alpha\sin{x\over N}\Big)^N=\nonumber\\
&&\Big(1+(\alpha^2-1)\sin^2{x\over N}\Big)^{N\over2}e^{iN\arctan(\alpha \tan{x\over N})}\eeqa

\bigskip
\noindent
First, let us consider the absolute value of $f$. We have
 \beqa 1&\leq& |f(x)|=\Big(1+(\alpha^2-1)\sin^2{x\over N}\Big)^{N\over2}\nonumber\\&\leq&\Big(1+(\alpha^2-1){{x^2}\over {N^2}}\Big)^{N\over2}\nonumber\\&\leq&\Big(1+(\alpha^2-1){1\over {N^{1+2\epsilon}}}\Big)^{N\over2}\rightarrow 1\label{absolute value limit}\eeqa
where in the last inequality we have used $|x|\leq {\cal O}(N^{1/2-\epsilon})$ and where the final limit is standard.

\bigskip
\noindent
Let us consider now the phase. For the approximation below all we require is ${x\over N}<<1$, which can be fulfilled for $|x|=\mu N$ where $\mu<<1$ is an arbitrary fixed constant. Then, using first order approximation 
\beq N\arctan(\alpha \tan{x\over N})\approx N \arctan(\alpha {x\over N})\approx N \alpha {x\over N}=\alpha x.\eeq

\noindent
$\square$ {\it QED}

\bigskip
\noindent Note that the region where $f(x)$ looks like a plane wave of wavenumber $\alpha$ is of order $O(N^{1/2-\epsilon})$ and that the limitation to this range follows from the behaviour of the absolute value of $f$. Indeed, for $|x|$ of order $O(N^{1/2})$ the absolute value of $f$ starts increasing; in particular, for $|x|=N^{1/2}$ we get $|f(x)|\rightarrow e^{{\alpha^2-1}\over 2}$ as we can readily see by using this value in (\ref{absolute value limit}). On the other hand, the phase continues to superoscillate on a much larger region, of order $N$. 

\section*{Supplementary Information II}

For simplicity, the preparation evolution described in eq (20) of the main text, (given below again for convenience)
\beq  |0\ra_p|\Phi\ra_P\rightarrow |\chi\ra_{p, P}=\sum_{m=-N}^N c_m|m\ra_p|\Phi -m\ra_{P}\label{entangled preparation 2}\eeq
has been described only as it acts on $ |0\ra_p|\Phi\ra_P$, our particular initial state of interest. Here we would like to show that this evolution can be implemented by a unitary that conserves angular momentum. For this we have to define the action of evolution operator on all the states in the Hilbert space of the particle and preparer. 

\bigskip
\noindent 
Suppose that we have a transformation $\hU$ such that
\beq  \hU |0\ra_p|k\ra_P =\sum_{m=-N}^N c_m|m\ra_p|k-m\ra_{P}\label{preparation unitary 1}\eeq 
with $-\infty<k<\infty$. 

\bigskip
\noindent 
Then, for any initial wavefunction $|{\tilde \Phi}\ra_P$ of the preparer we have
\beq  \hU |0\ra_p|{\tilde \Phi}\ra_P =\sum_{m=-N}^N c_m|m\ra_p|{\tilde \Phi}-m\ra_{P}\label{preparation unitary 2}\eeq 
as we can easily see if we insert in (\ref{preparation unitary 2}) the angular momentum decomposition of $|{\tilde \Phi}\ra_P$  and use (\ref{preparation unitary 1}). In other words, having a transformation $\hU$ that fulfils (\ref{preparation unitary 1}) is sufficient for implementing the transformation (\ref{entangled preparation 2}) that we desire.

\bigskip
\noindent 
It is convenient to write the transformation (\ref{preparation unitary 1}) in total and relative  angular momentum variables, $\hL_t=\hL_p+\hL_P$ and $\hL_r=\hL_p-\hL_P$.  In these new variables the transformation reads
 \beq  \hU |-k\ra_r|k\ra_t =\sum_{m=-N}^N c_m|m-k\ra_r|k\ra_t=|\xi_{-k,k}\ra_r|k\ra_t\label{preparation unitary 3}\eeq 
 where $|\xi_{-k,k}\ra_r=\sum_{m=-N}^N c_m|m-k\ra_r$.

\bigskip
\noindent We can now extend the transformation  (\ref{preparation unitary 3}) to a full, angular momentum conserving, unitary. For this we have to extend it to all possible initial states $|n\ra_r|k\ra_t$ with $-\infty<n, k<\infty$, i.e. to define its action on an entire basis of states. All we have to do is to define

 \beq  \hU |n\ra_r|k\ra_t =|\xi_{n,k}\ra_r|k\ra_t\label{preparation unitary 4}\eeq 
 with the states $|\xi_{n,k}\ra_r$ are arbitrary, except $|\xi_{n-k,k}\ra_r$ which is fixed by (\ref{preparation unitary 3}), and obey the orthogonality conditions
\beq _r\la\xi_{n,k}|\xi_{n',k}\ra_r=\delta_{n,n'}.\eeq

\bigskip
\noindent 
We can now see that for all the initial states $|0\ra_p|k\ra_P$ the transformation $\hU$ is angular momentum conserving and takes orthogonal states into orthogonal states, as required by a unitary.  Since for any fix total angular momentum $k$ a single state of relative momentum, namely  $|\xi_{-k,k}\ra_r$ is fixed by our desired transformation (\ref{entangled preparation 2}), we have plenty (infinite) of liberty to chose the states $|\xi_{n,k}\ra_r$ with $n\neq -k$ so that together with $|\xi_{-k,k}\ra_r$ they form a basis. Clearly then, the transformation  (\ref{preparation unitary 4}) is unitary (since it transforms the orthonormal basis $\{|n\ra_r|k\ra_t\}$ into the orthonormal basis $\{\xi_{n,k}\ra_r|k\ra_t\}$,) it is conserving the total angular momentum and  it implements on the state $ |0\ra_p|\Phi\ra_P$ the transformation we desire. 

\end{document}